\documentclass[aps,preprint,showpacs,groupedaddress,floatfix]{revtex4}
\usepackage{graphicx}
\usepackage{dcolumn}
\usepackage{bm}

\newcommand{\lp}{\left}
\newcommand{\rp}{\right}

\newcommand{\ket}[1]{\left| #1 \right\rangle}

\begin{document}

\bibliographystyle{prsty}

\title{Calculation of $\beta$-decay rates in a relativistic model with momentum-dependent self-energies}
\author{T. Marketin}
\author{D. Vretenar}
\affiliation{Physics Department, Faculty of Science, University of Zagreb, 
Croatia, and\\
Physik-Department der Technischen Universit\"at M\"unchen, 
D-85748 Garching,
Germany}
\author{P. Ring}
\affiliation{Physik-Department der Technischen Universit\"at M\"unchen, 
D-85748 Garching,
Germany}
\date{\today}

\begin{abstract}
The relativistic proton-neutron quasiparticle random phase approximation (PN-RQRPA) 
is applied in the calculation of $\beta$-decay half-lives of neutron-rich nuclei in the 
$Z\approx 28$ and $Z\approx 50$ regions. The study is based on the relativistic 
Hartree-Bogoliubov calculation of nuclear ground-states, using effective Lagrangians with 
density-dependent meson-nucleon couplings, and also extended by the inclusion of couplings 
between the isoscalar meson fields and the derivatives of the nucleon fields. This
leads to a linear momentum dependence of the scalar and vector nucleon self-energies.
The residual QRPA interaction in the particle-hole channel includes  the $\pi + \rho$ 
exchange plus a Landau-Migdal term.  The finite-range Gogny interaction is employed 
in the $T=1$ pairing channel, and the model also includes a proton-neutron particle-particle 
interaction. The results are compared with available data, and it is shown that an extension of the 
standard relativistic mean-field framework to include momentum-dependent nucleon self-energies 
naturally leads to an enhancement of the effective (Landau) nucleon mass, 
and thus to an improved PN-QRPA description of $\beta^-$-decay rates.
\end{abstract}

\maketitle

Weak-interaction processes in exotic nuclei far from stability play an important role in stellar 
explosive events. In particular,  $\beta$-decay rates of very neutron-rich nuclei set the time scale 
of the $r$-process nucleosynthesis, i.e. the multiple neutron capture process which determines the 
synthesis of nearly half of the nuclei heavier than Fe. Since the vast majority of nuclides which lie on 
the path of the $r$-process will not be experimentally accessible in the foreseeable future, it is 
important to develop microscopic nuclear structure models that can provide accurate predictions of 
weak-interaction rates of thousands of nuclei with large neutron to proton asymmetry. There are 
basically two microscopic approaches that can be employed in large-scale calculations of 
$\beta$-decay rates: the interacting shell model and the quasiparticle random phase approximation 
(QRPA).  While the advantage of using the shell model is the ability to take into account the 
detailed structure of the $\beta$-strength function \cite{LM.03}, the QRPA approach is based 
on global effective interactions and provides a systematic description of $\beta$-decay properties 
of arbitrarily heavy nuclei along the $r$-process path \cite{Bor.06}. In a recent review of modern 
QRPA calculations of $\beta$-decay rates for astrophysical applications \cite{Bor.06}, Borzov has 
emphasized the importance of performing calculations based on self-consistent mean-field 
models, rather than on empirical mean-field potentials, e.g. the Woods-Saxon potential. In a 
self-consistent framework both the nuclear ground states, i.e. the masses which determine the 
possible $r$-process path, and the corresponding 
$\beta$-decay properties are calculated from the same energy 
density functional or effective nuclear interaction. This approach ensures the consistency of the 
nuclear structure input for astrophysical modeling, and allows reliable extrapolations of the nuclear 
spin-isospin response to regions of very neutron-rich nuclei. 

The fully consistent proton-neutron (PN) relativistic QRPA \cite{VPNR.03,Paa.04} has recently been 
employed in the calculation of $\beta$-decay half-lives of neutron-rich nuclei in the 
N$\approx$50 and N$\approx$82 regions \cite{NMV.05}. The model is based on the 
relativistic QRPA \cite{Paa.03}, formulated in the canonical basis of the relativistic Hartree-Bogoliubov 
(RHB) framework \cite{VALR.05}.  The RHB+RQRPA model is fully self-consistent.
For the interaction in the particle-hole channel modern effective Lagrangians with
density-dependent meson-nucleon couplings are used, and pairing correlations are
described by the pairing part of the finite range Gogny interaction. Both in
the particle-hole ($ph$) and particle-particle ($pp$) channels, the same interactions
are used in the RHB equations which determine the nuclear ground-state, and in the
matrix equations of the RQRPA. This is important because the energy
weighted sum rules are only satisfied if the pairing interaction is
consistently included both in the static RHB and in the dynamical RQRPA
calculations. In both channels the same strength parameters of the 
interactions are used in the RHB and RQRPA calculations. The formulation of the RHB+RQRPA 
model in the canonical quasiparticle basis enables the description of weakly-bound neutron-rich 
nuclei far from stability, because this basis diagonalizes the density matrix and includes both the 
bound states and the positive-energy single-particle continuum \cite{Paa.03}. 

In the corresponding proton-neutron RQRPA \cite{VPNR.03,Paa.04} the spin-isospin-dependent 
interaction terms are generated by the $\pi$- and $\rho$-meson exchange.
Although the direct one-pion contribution to the nuclear ground state vanishes at the mean-field 
level because of parity conservation, it must be included in the calculation of spin-isospin excitations.
In addition, the derivative type of the pion-nucleon coupling necessitates the inclusion of the zero-range
Landau-Migdal term, which accounts for the contact part of the nucleon-nucleon interaction, 
with the strength parameter $g^{\prime}$ adjusted to reproduce experimental data on the excitation
energies of Gamow-Teller resonances (GTR). The model also includes the $T=0$ proton-neutron 
pairing interaction: a short-range repulsive Gaussian function combined with a
weaker longer-range attractive Gaussian \cite{Eng.99}. In general the calculated $\beta$-decay 
half-lives are very sensitive to the strength of the $T=0$ pairing which, in the case of $\beta^-$-decay, 
enhances the Gamow-Teller strength in the $Q_{\beta^-}$ energy window.

Standard relativistic mean-field models are based on the static approximation, i.e. the nucleon 
self-energy is real, local and energy-independent. Consequently, these models describe correctly 
the ground-state properties and the sequence of single-particle levels in finite nuclei, but not the 
level density around the Fermi surface. The reason is the low effective nucleon mass $m^*$ which, 
in the relativistic framework, is also related to the Dirac mass $m_D = m + S(\bm{r})$, 
where $m$ is the bare nucleon mass and $S(\bm{r})$ denotes the scalar nucleon self-energy, 
and thus constrained by the empirical spin-orbit energy splittings. The difference between the
vector and scalar nucleon self-energies determines the spin-orbit potential, whereas their 
sum defines the effective single-nucleon potential, and is constrained by the nuclear matter
binding energy at saturation density. The energy spacings between spin-orbit partner states in 
finite nuclei, and the nuclear matter binding and saturation, place the following constraints on the 
values of the Dirac mass and the nucleon effective mass: $0.55 m \le m_D \le 0.6 m$,  $0.64 
m \le m^* \le 0.67 m$, respectively. These values have been used in most standard relativistic 
mean-field effective interactions. However, when these interactions are used in the calculation 
of $\beta^-$ decay rates, the resulting half-lives are usually more than an order of magnitude 
longer than the empirical values. This is because the low effective nucleon mass implies 
a low density of states around the Fermi surface, and therefore in a self-consistent relativistic 
QRPA calculation of $\beta$-decay the transition energies will be low, resulting in long lifetimes. 
In order to reproduce the empirical half-lives, it is thus necessary to employ relativistic effective 
interactions with higher values of the nucleon effective mass. We note that in the case of 
non-relativistic global effective interactions such as, for instance, Skyrme-type interactions, 
calculation of ground-state properties and excitation energies of quadrupole giant resonances 
have shown that a realistic choice for the nucleon effective mass is in the 
interval $m^*/m = 0.8\pm 0.1$~\cite{Rei.99,Cha.97}.

In Ref.~\cite{NMV.05} we have used the RHB+RQRPA model to calculate $\beta$-decay half-lives 
of neutron-rich nuclei in the N$\approx$50 and N$\approx$82 regions. Starting from the standard 
density-dependent effective interaction DD-ME1 \cite{NVFR.02} ($m_D = 0.58\,m$, $m^*=0.66\,m$), 
a new effective interaction was adjusted with higher values for the Dirac mass and the nucleon 
effective mass: $m_D = 0.67\,m$, $m^* = 0.76\, m$. However, a standard RMF interaction with 
such a high value of the Dirac mass would systematically underestimate the empirical spin-orbit 
splittings in finite nuclei. To compensate the reduction of the effective spin-orbit potential caused 
by the increase of the Dirac mass, the DD-ME1 interaction was further extended by including 
an additional interaction term:  the tensor coupling of the $\omega$-meson to the nucleon. 
The resulting interaction was used in the relativistic Hartree-Bogoliubov calculation of nuclear ground 
states. With the Gogny D1S interaction in the $T=1$ pairing channel, and also including the 
$T=0$ particle-particle interaction in the PN-QRPA, it was possible 
on one hand to reproduce the empirical values of the energy spacings between spin-orbit 
partner states in spherical nuclei, and on the other hand the calculated $\beta$-decay half-lives 
were in reasonable agreement with the experimental data for the Fe, Zn, Cd, and Te isotopic chains.

With the model developed in Ref.~\cite{NMV.05} the problems of the low effective mass and long 
$\beta$-decay half-lives were solved on an ad hoc basis. The effective interaction was adjusted 
with the particular purpose of increasing the effective nucleon mass, and the resulting problem 
of the reduction of the effective spin-orbit potential was solved by the inclusion of an additional 
interaction term. A much better solution is provided by the recently introduced relativistic mean-field 
model with momentum-dependent nucleon self-energies \cite{Typ.03,Typ.05}. In this model
the standard effective Lagrangian with density-dependent meson-nucleon coupling vertices is 
extended by including a particular form of the couplings between the isoscalar meson fields and the 
derivatives of the nucleon fields. This leads to a linear momentum dependence of the scalar and 
vector self-energies in the Dirac equation for the in-medium nucleon. Even though the extension 
of the standard mean-field framework is phenomenological, it is nevertheless based on Dirac-Brueckner 
calculations of in-medium nucleon self-energies, and consistent with the relativistic optical potential in 
nuclear matter, extracted from elastic proton-nucleus scattering data. In the extended model it is 
possible to increase the effective nucleon mass, while keeping a small Dirac mass which is 
required to reproduce the empirical strength of the effective spin-orbit potential. 

In the very recent work of Ref.~\cite{Typ.05}, in particular, an improved Lagrangian density of 
the model with density-dependent and derivative couplings (D$^3$C) has been introduced. 
The parameters of the coupling functions were adjusted to ground-state properties of eight 
doubly-magic spherical nuclei, and the results for nuclear matter, neutron matter, and finite
nuclei were compared to those obtained with conventional RMF models. It was shown that 
the new effective interaction improves the description of binding energies, nuclear shapes 
and spin-orbit splittings of single-particle levels. More important, it was possible to increase 
the effective nucleon mass ($m^* = 0.71\,m$) and, correspondingly, the density of 
single-nucleon levels close to the Fermi surface as compared to standard RMF models. 
At the same time the Dirac mass was kept at the small value $m_D = 0.54m$, which 
ensures that the model reproduces the empirical spin-orbit splittings. The momentum 
dependence of the nucleon self-energies provides also a correct description of the 
empirical Schroedinger-equivalent central optical potential.

In this work we employ the model with density-dependent and derivative couplings (D$^3$C) 
of Ref.~\cite{Typ.05} in the calculation of $\beta$-decay rates of neutron-rich nuclei in several 
isotopic chains in the $Z \approx 28$ and $Z \approx 50$ regions. The results of fully consistent 
RHB plus proton-neutron QRPA will be compared with those obtained with the standard 
density-dependent RMF interaction DD-ME1 and, in addition, with a new effective 
interaction based on the D$^3$C model, but with an even higher value of the effective 
nucleon mass. We will analyze the dependence of the $\beta$-decay half-lives on the 
choice of the effective particle-hole interaction, and the strength of the $T=0$ pairing 
interaction. 

The functional forms of the density dependence of the $\sigma$, $\omega$ and $\rho$ 
meson-nucleon couplings are identical for the conventional DD-ME1 effective interaction 
and the D$^3$C model. The latter includes momentum-dependent isoscalar scalar and 
vector self-energies, and thus contains two additional coupling functions $\Gamma_{S}$ 
and $\Gamma_{V}$. In Ref.~\cite{Typ.05} these have been parametrized with the 
following functional form:
\begin{equation}
\Gamma_{i}(x) = \Gamma_{i}(\rho_{ref}) x^{-a_{i}} \quad \textrm{for} \quad i=S, V \; ,
\end{equation}
where $x = \rho_v / \rho_{ref}$, $\rho_v$ is the vector density, and the reference density 
$\rho_{ref}$ corresponds to the vector density determined at the saturation point of 
symmetric nuclear matter. In the parameterization of  Ref.~\cite{Typ.05} $a_S = a_V = 1$, 
and we will retain these values in the following calculation. The parameters 
$ \Gamma_{S}(\rho_{ref})$ and  $\Gamma_{V}(\rho_{ref})$ have been constrained by 
the requirement that the resulting optical potential in symmetric nuclear matter at 
saturation density has the value 50 MeV at a nucleon energy of 1 GeV. In total there are 
10 adjustable parameters in the D$^3$C model, compared to eight for the standard 
density-dependent RMF models, e.g. the DD-ME1 parameterization.

The effective nucleon mass of the D$^3$C model is $m^* = 0.71\, m$, compared to 
$m^* = 0.66 \,m$ for DD-ME1. In addition, starting from  D$^3$C, for the purpose 
of calculating $\beta$-decay rates we have adjusted a
new parameterization with  $m^* = 0.79\, m$, which is much closer to the effective 
masses used in non-relativistic Skyrme effective interactions \cite{Rei.99,Cha.97}. The new 
effective interaction which, for simplicity we denote  D$^3$C{\large *}, has been adjusted
following the original procedure of Ref.~\cite{Typ.05}, with an additional constraint on 
the effective nucleon mass.  Even though we have tried to increase the effective mass 
as much as possible, $m^* = 0.79\, m$ is the highest value 
for which a realistic description of nuclear matter and finite 
nuclei is still possible, and the quality of the calculated 
nuclear matter equation of state and of ground-state properties of spherical 
nuclei is comparable to that of the DD-ME1 and D$^3$C interactions. The three 
interactions are compared in Table~\ref{TabA}, where we include the characteristics
of the corresponding nuclear matter equations of state at saturation point: 
the saturation density $\varrho_{\rm sat}$, 
the binding energy per particle $a_{V}$, the symmetry energy $a_{4}$, 
the nuclear matter compression modulus  $K_{\infty}$, 
the Dirac mass $m_{D}$, and the effective (Landau) mass $m^{*}$.
In addition, for the two interactions with 
energy-dependent single-nucleon potentials, we compare the values of 
$ \Gamma_{S}(\rho_{ref})$ and  $\Gamma_{V}(\rho_{ref})$. We notice a 
pronounced increase of the strength of the scalar field. This is, however, compensated 
by the corresponding decrease of the strength of the vector coupling, so that the 
difference $ \Gamma_{V}(\rho_{ref}) - \Gamma_{S}(\rho_{ref})$ is practically the same 
for D$^3$C and D$^3$C{\large *}. For both interactions the optical potential at 1 GeV 
nucleon energy  has been constrained to 50 MeV. With the increase of the effective 
nucleon mass from DD-ME1 to D$^3$C and D$^3$C$^*$, we also note the 
corresponding decrease of the nuclear matter compression modulus $K_{\infty}$ . 
This  correlation between $K_{\infty}$ and $m^*$ is also well known in 
non-relativistic Skyrme effective interactions \cite{Cha.97}.

\begin{table}
\caption{\label{TabA}
Properties of symmetric nuclear matter at saturation density calculated 
with the models DD-ME1, D$^{3}$C, and D$^3$C{\large *}.
}
%\begin{ruledtabular}
\begin{tabular}{cccc}
 & DD-ME1 & D$^{3}$C & D$^3$C{\large *} \\
 \hline
 $\varrho_{\rm sat}$ [fm${}^{-3}$]  & 0.152 & 0.151& 0.152 \\
 $a_{V}$ [MeV]                     & -16.20 & -15.98 & -16.30 \\
 $a_{4}$ [MeV]                     & 33.1 & 31.9 & 33.0\\
 $K_{\infty}$ [MeV]                  & 244.5 & 232.5 & 224.9\\
  $m_{D}/m$                        & 0.58 & 0.54 & 0.57\\
 $m^{*}/m$                     & 0.66 & 0.71 & 0.79 \\
 $\Gamma_{S}$  & 0.0 & -21.632 & -146.089 \\
$\Gamma_{V}$  & 0.0 & 302.188 &  180.889 \\
 \end{tabular}
%\end{ruledtabular}
\end{table}

\begin{figure}
\centerline{ \includegraphics[scale=0.7]{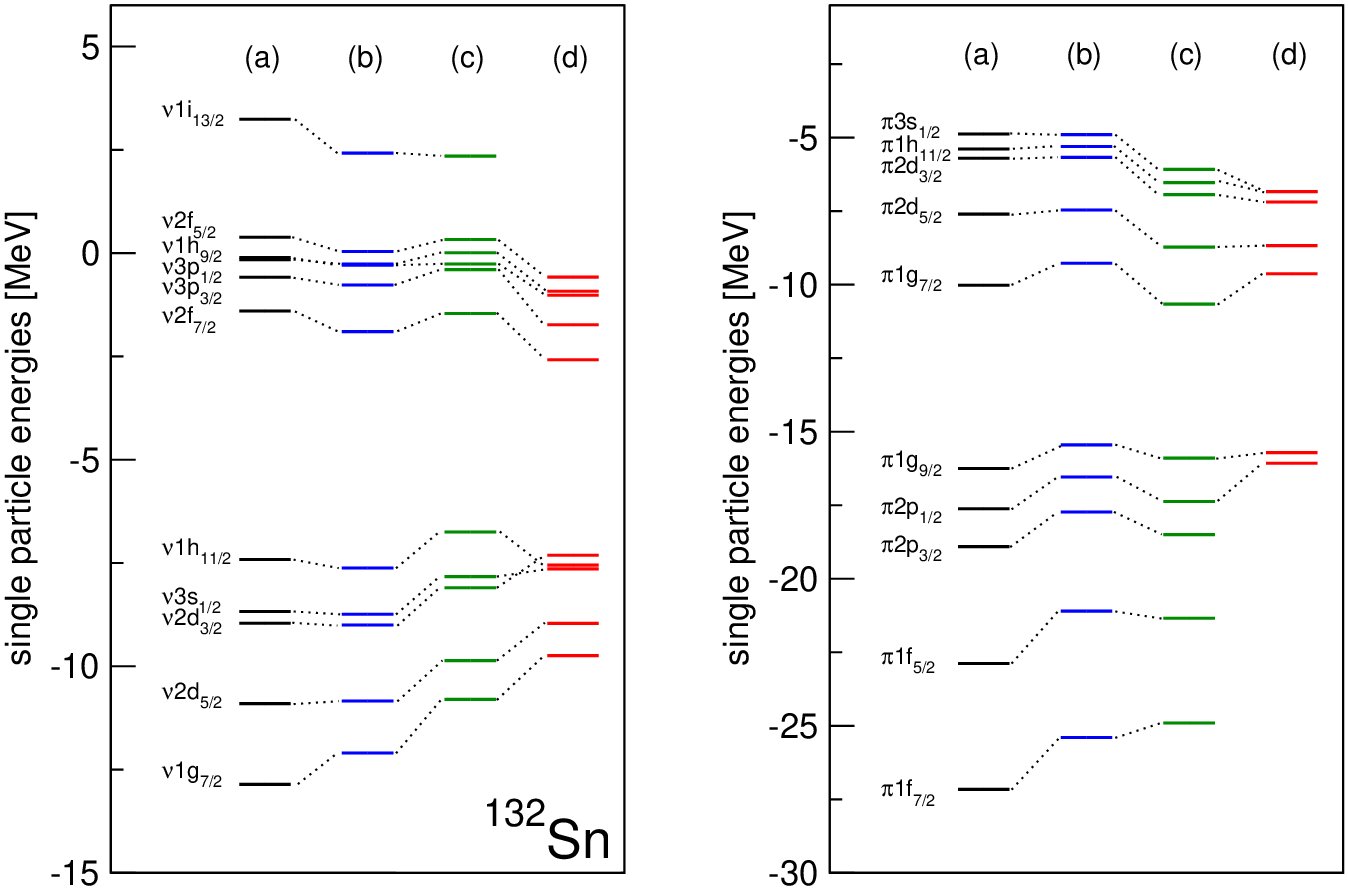} }
\caption{(Color online) Neutron (left panel) and proton (right panel) single-particle levels in $^{132}\textrm{Sn}$  
calculated with the DD-ME1 (a), D$^{3}$C (b) and D$^3$C{\large *} (c) interactions, 
compared to experimental levels (d) \protect\cite{Isa.02}.}
\label{Fig1}
\end{figure}

In Fig.~\ref{Fig1} we display the neutron and proton single-particle levels in  $^{132}$Sn
calculated in the relativistic mean-field model with the DD-ME1,  D$^{3}$C, and D$^3$C{\large *} 
effective interactions, in comparison with available data for the levels close to the Fermi 
surface \cite{Isa.02}. Compared to the DD-ME1 interaction, the enhancement 
of the effective mass in D$^{3}$C and D$^3$C{\large *} results in the increase of the density
of states around the Fermi surface, and the calculated spectra are in much better 
agreement with the empirical energy spacings.

In the next step  the three effective interactions have been tested and compared in RHB 
plus proton-neutron relativistic QRPA calculations 
of $\beta$-decay half-lives for the isotopic chains: Fe, Ni, Zn, Cd, Sn and Te. 
The nuclear ground-states have been calculated in the RHB model with the DD-ME1, 
D$^{3}$C, and D$^3$C{\large *} effective interactions in the particle-hole channel, and the 
pairing part of the Gogny force,
\begin{equation}
V^{pp}(1,2)~=~\sum_{i=1,2}e^{-((\mathbf{r}_{1}-\mathbf{r}_{2})/{\mu _{i}}%
)^{2}}\,(W_{i}~+~B_{i}P^{\sigma }-H_{i}P^{\tau }-M_{i}P^{\sigma }P^{\tau })
\label{Gogny}
\end{equation}
in the particle-particle channel, with the set D1S \cite{BGG.91} for the
 parameters $\mu _{i}$, $W_{i}$, $B_{i} $, $H_{i}$ and $M_{i}$ $(i=1,2)$. 
 This force has been very carefully
adjusted to pairing properties of finite nuclei all over the periodic
table. In particular, the basic advantage of the Gogny force is the finite
range, which automatically guarantees a proper cut-off in momentum space.
In the following calculations we have also used the Gogny interaction
in the $T=1$ $pp$-channel of the PN-RQRPA. 

The RHB ground-state solution determines the single-nucleon canonical basis, i.e. 
the configuration space in which the matrix equations of the relativistic QRPA are 
expressed (see Refs.~\cite{Paa.03,Paa.04} for a detailed presentation of the formalism). 
The particle-hole residual interaction of the PN-RQRPA is derived from the 
following Lagrangian density:
\begin{equation}
\mathcal{L}_{\pi + \rho}^{int} = 
      - g_\rho \bar{\psi}\gamma^{\mu}\vec{\rho}_\mu \vec{\tau} \psi 
      - \frac{f_\pi}{m_\pi}\bar{\psi}\gamma_5\gamma^{\mu}\partial_{\mu}
        \vec{\pi}\vec{\tau} \psi \; . 
\label{lagrres} 
\end{equation}
The coupling between the $\rho$-meson and the nucleon is already contained in 
the RHB effective Lagrangian, and the same interaction is consistently used in 
the isovector channel of the QRPA. The direct one-pion contribution to the 
ground-state RHB solution vanishes because of parity-conservation, but it 
must be included in the calculation of the Gamow-Teller strength.  
For the pseudovector pion-nucleon coupling we have used the standard values:
\begin{equation}
m_{\pi}=138.0~{\rm MeV}~~~~\;\;\;\;\frac{\;f_{\pi}^{2}}{4\pi}=0.08\;.
\end{equation}
In addition, the zero-range Landau-Migdal term accounts for the contact part of the 
isovector channel of the nucleon-nucleon interaction
\begin{equation}
V_{\delta \pi} = g' \lp( \frac{f_{\pi}}{m_{\pi}} \rp)^{2} \vec{\tau}_{1} 
\vec{\tau}_{2} \bm{\Sigma_{1}}\cdot \bm{\Sigma_{2}} \delta\lp( \bm{r}_{1}-\bm{r}_{2} \rp) \; .
\end{equation}
For each effective interaction, the strength parameter $g^\prime$ is adjusted to reproduce 
the excitation energy of the Gamow-Teller resonance in $^{208}$Pb. In the present 
calculation these values are:  $g^\prime = 0.55, 0.54$ and $0.76$, for DD-ME1, 
D$^{3}$C, and D$^3$C{\large *}, respectively. 

Finally, the proton-neutron QRPA interaction is completely determined by the choice of 
the $T=0$ pairing interaction \cite{Eng.99}:
\begin{equation}
\label{eq2}
V_{12}
= - V_0 \sum_{j=1}^2 g_j \; {\rm e}^{-r_{12}^2/\mu_j^2} \;
    \hat\Pi_{S=1,T=0}
\quad ,
\label{pn-pair}
\end{equation}
where $\hat\Pi_{S=1,T=0}$ projects onto states with $S=1$ and $T=0$.  
The ranges $\mu_1$=1.2\,fm and $\mu_2$=0.7\,fm of the two Gaussians are the same
as for the Gogny interaction Eq. (\ref{Gogny}), and the relative strengths $g_1 =1$ and
$g_2 = -2$ are adjusted so that the force is repulsive at small distances.  The only
remaining free parameter is $V_0$, the overall strength.

The half-life of the $\beta^-$-decay of an even-even nucleus in the allowed 
Gamow-Teller approximation is calculated from the following expression:
\begin{equation}
\label{halflife}
\frac{1}{T_{1/2}}=\sum_m{\lambda_{if}^m} = D^{-1}g_A^2\sum_m
  \int{dE_e \lp| \sum_{pn} <1^+_m||\bm{\sigma}\tau_-||0^+>\rp|^2 
   \frac{dn_m}{dE_e}}\;,
\end{equation}
where $D=6163.4\pm 3.8$ s \cite{BG.00}.
$\ket{0^+}$ denotes the ground state of the parent nucleus, and
$\ket{1^+_m}$ is a state of the daughter nucleus.
The sum runs over all final states with an excitation energy smaller than the
$Q_{\beta^-}$ value. In order to account for the universal quenching of
the Gamow-Teller strength function, we have used the effective weak axial nucleon
coupling constant $g_A=1$, instead of $g_A=1.26$~\cite{BM.75}. The kinematic
factor in Eq.~(\ref{halflife}) can be written as
\begin{equation}
\label{kinematic}
\frac{dn_m}{dE_e}=E_e\sqrt{E_e^2-m_e^2}(\omega -E_e)^2 F(Z,A,E_e)\;,
\end{equation}
where $\omega$ denotes the energy difference between the initial 
and the final state. The Fermi function $F(Z,A,E_e)$ corrects the phase-space
factor for the nuclear charge and finite nuclear size effects~\cite{KR.65}. 

\begin{figure}
\centerline{ \includegraphics[scale=0.7]{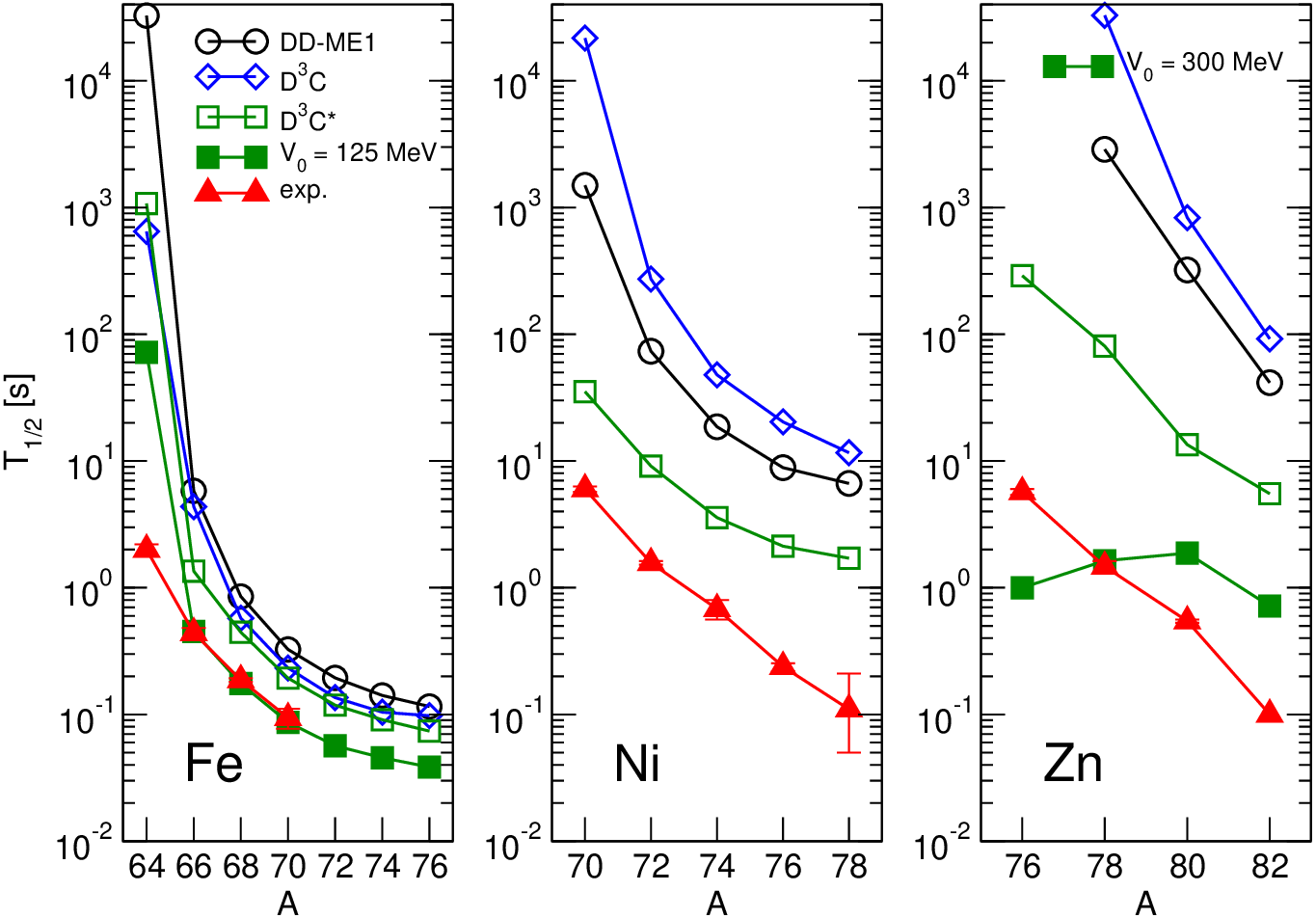} }
\caption{(Color online) $\beta$-decay half-lives of Fe (left panel), Ni (middle panel), and Zn (right panel) nuclei,
calculated with the DD-ME1, D$^{3}$C, and D$^3$C{\large *} effective interactions, compared with 
the experimental values \protect\cite{NUDAT}. Open symbols correspond to PN-QRPA values 
calculated without the inclusion of the $T=0$ pairing interaction. 
The filled squares are half-lives calculated with 
the D$^3$C{\large *} interaction and $T=0$ pairing, with the strength parameter $V_0 = 125$ MeV for 
Fe, and  $V_0 = 300$ MeV for Zn isotopes.}
\label{Fig2}
\end{figure}

In Figure~\ref{Fig2} we display the $\beta^-$-decay half-lives of iron, nickel and zinc isotopes 
calculated with the DD-ME1, D$^{3}$C, and D$^3$C{\large *}, and compare them with the experimental 
values taken from NUDAT database \cite{NUDAT}. The data for $^{76}$Ni and $^{78}$Ni are from 
Ref.~\cite{Hos.05}.  Open symbols correspond to values calculated 
without the inclusion of  $T=0$ pairing. Since the $\beta^-$-decay rates are generally very sensitive 
to the proton-neutron pairing, and its strength is usually adjusted separately for each isotopic 
chain, we will first discuss the results obtained without the $T=0$ pairing interaction. 
For all three isotopic chains, the shortest half-lives are obtained with the interaction with the 
highest effective mass, i.e. D$^3$C{\large *}, even though these are still far from the experimental values. 
For the Fe nuclei all three interactions give similar results, whereas more pronounced differences are 
found for the Ni and Zn isotopic chains. In the two latter cases  similar results are obtained with 
DD-ME1 and D$^{3}$C and, in fact, longer half-lifes are predicted by D$^{3}$C, even though 
it has a higher effective nucleon mass. Much shorter half-lives for the Ni and Zn nuclei are calculated 
with the  D$^3$C{\large *} effective interaction. The origin of these large differences in the calculated rates 
can be understood from Table~\ref{TabB}, where we list the transition energies for the strongest 
transition in the Zn isotopes with $76 \leq A \leq 82$:  $\nu 2p_{1/2} \to \pi 2p_{3/2}$. We 
note that the transition energies for the DD-ME1 and D$^{3}$C interactions are comparable and,
in particular, those calculated with  DD-ME1 are slightly larger, resulting in faster $\beta^-$-decay rates. 
Both interaction predict a $\beta$-stable $^{76}\textrm{Zn}$.
On the other hand, the transition energies predicted by the interaction D$^3$C{\large *} are much larger 
and, correspondingly, the calculated half-lives are at least an order of magnitude shorter.

\begin {table}[h]
\setlength{\tabcolsep}{5.0\tabcolsep}
\caption{Transition energies (in MeV) for the strongest transition in the 
Zn isotopes: $\nu 2p_{1/2} \to \pi 2p_{3/2}$. }
\begin {center}
\begin {tabular}{c c c c}
\hline
\hline
 & DD-ME1 & D$^{3}$C & D$^3$C{\large *} \\
\hline
$^{76}\textrm{Zn}$ & 0.15 & -0.05 & 1.74 \\
$^{78}\textrm{Zn}$ & 0.93 & 0.72 & 2.65 \\
$^{80}\textrm{Zn}$ & 2.01 & 1.80 & 3.69 \\
$^{82}\textrm{Zn}$ & 2.69 & 2.51 & 4.58 \\
\hline
\hline
\end {tabular}
\end{center}
\label{TabB}
\end{table}

\begin{figure}
\centerline{ \includegraphics[scale=0.7]{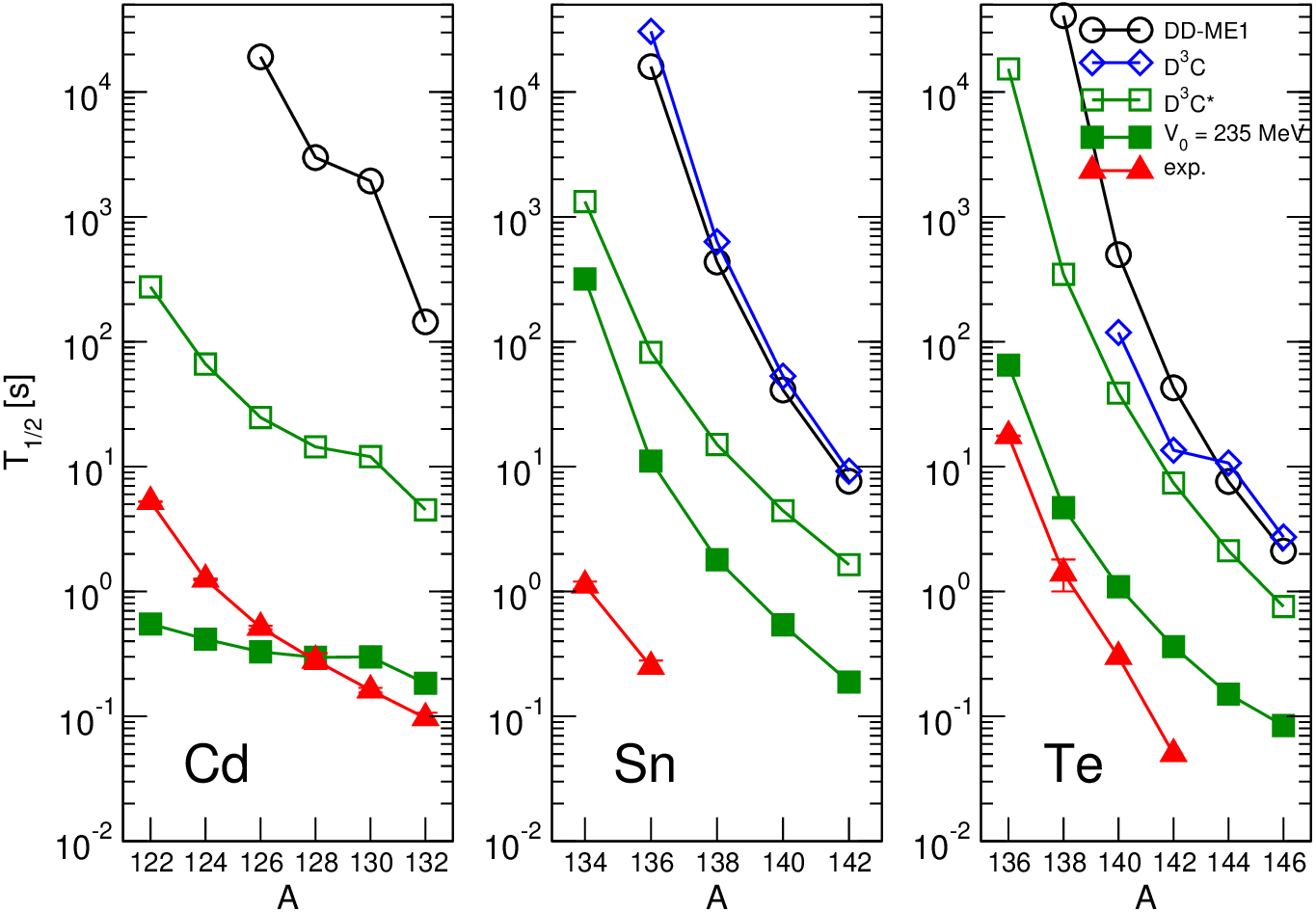} }
\caption{(Color online) Same as in Fig.~\ref{Fig2}, but for the Cd (left panel), Sn (middle panel), and Te 
(right panel) isotopic chains. 
For the D$^3$C{\large *} effective interaction, in all three isotopic chains the strength of the 
$T=0$ pairing interaction is $V_0 = 235$ MeV (filled squares).}
\label{Fig3}
\end{figure}

A similar situation is found in the neutron-rich nuclei in the $Z \approx 50$ region. 
The calculated half-lives of Cd, Sn and Te nuclei are plotted in Fig.~\ref{Fig3}, in comparison 
with available data \cite{NUDAT}. The Cd isotopes, in particular, are calculated as $\beta$-stable 
with the D$^{3}$C interaction, because the predicted transition energies are smaller than 
the electron rest mass. Much better results are obtained with the modified interaction 
D$^3$C{\large *}, which clearly reproduces the isotopic trend of the half-lives of neutron-rich 
Cd nuclei. DD-ME1 and D$^{3}$C produce almost identical results for Sn and Te nuclei. 
Shorter half-lives, especially for Sn, are calculated with D$^3$C{\large *}, but these are still 
orders of magnitude from the experimental values. It appears that all three interactions 
reproduce the isotopic trend in the Te chain.

We have considered the effect of the $T=0$ pairing interaction on the calculated 
$\beta$-decay half-lives only for the D$^3$C{\large *} effective interaction which, with the 
effective nucleon mass $m^*/m = 0.79$ comparable to those of non-relativistic 
effective interactions, gives the shortest half-lives.  Even without the
inclusion of the proton-neutron $pp$ interaction, for the Fe nuclei the 
calculated half-lives are already close to the experimental values, except for 
$^{64}$Fe (see Fig.~\ref{Fig2}).  By adjusting the value of the strength parameter
of the $T=0$ pairing to $V_0 = 125$ MeV, the PN-QRPA calculation reproduces 
the $\beta$-decay half-lives of $^{66}$Fe, $^{68}$Fe and $^{70}$Fe (filled squares). 
In the case of Ni isotopes the $T=0$ interaction in the $pp$-channel is not effective 
because of the $Z=28$ and $N=40$ closures \cite{Eng.99,NMV.05}. 
The $\pi 1f_{7/2}$ orbit is  completely occupied, and the transition $\nu 1f_{5/2} \to \pi 1f_{7/2}$ 
is thus blocked. The $T=0$  pairing could only have an effect on the $\nu 1g_{9/2} \to \pi 1g_{9/2}$
transition, but the $\pi 1g_{9/2}$ orbital is located high above the Fermi surface. Thus the  
$T=0$ $pp$ interaction cannot shift the low-energy GT strength and enhance the $\beta$-decay 
rates. Even using the D$^3$C{\large *} interaction, the calculated half-lives are an order of magnitude 
longer than the experimental values. 

The principal advantage of the self-consistent approach to the modeling of $\beta$-decay rates is 
the use of universal (A independent) interactions in the $ph$-channel and, in many cases including 
the model used in this work, in the $T=1$ $pp$-channel. Unfortunately, this is not possible in the 
$T=0$ $pp$-channel, for which no information is contained in the ground-state data. The strength 
of this interaction is adjusted separately for each 
isotopic chain or, in the best case, a single value of the strength can be used in a certain 
mass region \cite{Eng.99,NMV.05}. It is especially difficult to keep the same strength of the 
$T=0$ pairing when crossing a closed shell. Thus in going from the Fe to the Zn isotopic chain 
we had to increase the strength parameter $V_0$ by more than a factor two. The value 
$V_0 = 300$ MeV has been adjusted to reproduce the half-life of $^{78}$Zn (filled squares 
in the right panel of Fig. \ref{Fig2}) but, even though the calculated values are in qualitative 
agreement with the data, with the inclusion of the $T=0$ pairing the PN-QRPA results do 
not reproduce the isotopic dependence of the experimental half-lives. In other words, it 
was not possible to find a single value of the proton-neutron pairing strength that could 
reproduce the half-lives of neutron-rich Zn isotopes.

The filled squares in Fig.~\ref{Fig3} correspond to the half-lives calculated with the D$^3$C{\large *} 
effective interaction, the $\pi + \rho$ plus Landau-Migdal interaction in the $ph$-channel, 
the Gogny interaction Eq.~(\ref{Gogny}) in the $T=1$ $pp$-channel, and the $T=0$ pairing 
Eq.~(\ref{pn-pair}). The strength of the latter: $V_0=235$ MeV, 
has been adjusted to the half-life of $^{128}$Cd, and this value has been used for the Cd, Sn and 
Te isotopic chains. The effect of the $T=0$ pairing is especially pronounced for Cd and Te nuclei, 
and the results are in qualitative agreement with the available data, although the calculation does
not reproduce the isotopic trend for the Cd chain, and overestimates the half-lives of Te isotopes. 
On the other hand, for the proton closed-shell Sn nuclei the $T=0$ pairing interaction is much less
effective, and the calculated half-lives of $^{134}$Sn and $^{136}$Sn are two orders of magnitude
longer than the experimental values. Better results could be obtained, of course, by adjusting 
$V_0$ separately for each isotopic chain. 

The calculations performed in this work have shown that the extension of the standard relativistic 
mean-field framework to include momentum-dependent (energy-dependent in stationary systems) 
nucleon self-energies naturally leads to an enhancement of the effective (Landau) nucleon mass, 
and thus to an improved PN-QRPA description of $\beta^-$-decay rates. However, even though 
the momentum-dependent RMF model with density-dependent meson-nucleon couplings, 
adjusted here to $m^* = 0.79\, m$, predicts half-lives of neutron-rich medium-mass nuclei in 
qualitative agreement with data, the results are not as good as those obtained in the most 
advanced non-relativistic self-consistent density-functional plus continuum-QRPA 
calculations \cite{Bor.03,Bor.05,Bor.06}, or with the self-consistent HFB+QRPA 
model with Skyrme interactions of Ref.~\cite{Eng.99}. 
Namely, although we have been able to increase the effective mass of the 
interaction used in the RHB calculations of nuclear ground states to $m^* = 0.79\, m$, 
a value which is sufficient for the description of giant resonances  \cite{Rei.99,Cha.97}, 
the detailed description of the low-energy Gamow-Teller strength necessitates an even 
higher value of $m^*$. In fact, the effective mass of the Skyrme SkO' interaction 
used in Ref.~\cite{Eng.99} is $m^* = 0.9\, m$, whereas the continuum-QRPA calculations 
by Borzov are based on the Fayans phenomenological density functional with the 
bare nucleon mass, i.e.  $m^* =  m$ \cite{Bor.03,Bor.05,Bor.06}. However, it would 
be very difficult to further increase the effective nucleon mass in 
the framework of the model used in this work, i.e. on the nuclear matter level, 
without destroying the good agreement with empirical ground-state properties of 
finite nuclei. On the other hand, this would not even be the correct procedure because the 
additional enhancement of the effective nucleon mass is due to the coupling of 
single-nucleon levels to low-energy collective vibrational states, an effect which goes 
entirely beyond the mean-field approximation and is not included in the present model.
In principle, the effect of two- and three-phonon states on the weak-interaction rates could 
be taken into account by explicitly considering the coupling of 
single-quasiparticle states to phonons, and 
the resulting complex configurations would certainly lead to a redistribution of low-energy 
Gamow-Teller strength. Even though such extended (second) RPA approaches have been
routinely used for many years in the calculation of widths of isoscalar and isovector 
giant resonances, no systematic large-scale calculations of $\beta$-decay rates have 
been reported so far. We have therefore started to develop a new self-consistent model 
based on the recently introduced covariant theory of particle-vibration coupling \cite{LR.06}, 
and this framework will be applied in the calculation of $\beta$-decay half-lives of 
neutron-rich medium mass nuclei. 

In heavier nuclei, or in nuclei with an even higher neutron to proton asymmetry, in addition 
to allowed Gamow-Teller transitions, first-forbidden transitions must be taken into account 
in the calculation of $\beta$-decay half-lives. As it has been shown in recent studies by 
Borzov using the density-functional plus continuum-QRPA framework \cite{Bor.03,Bor.05,Bor.06},
the first-forbidden decays have a pronounced effect on 
the $\beta$-decay characteristics of $r$-process nuclei in the 
$Z\approx  28$, $N>50$; $Z\geq 50$, $N>82$; and $Z = 60 - 70$, $N \approx 126$ regions.
For studies of weak-interaction rates in  $r$-process nuclei very far from stability, it will therefore
be important to include first-forbidden transitions in the relativistic PN-QRPA model.

\bigskip
\bigskip
%\newpage
%=========================================================================
\leftline{\bf ACKNOWLEDGMENTS} 
\noindent
This work has been supported in part
by the Bundesministerium f\"ur Bildung und Forschung - project 06 MT
246, by the Gesellschaft f\"ur Schwerionenforschung GSI - project
TM-RIN.  T. Marketin and D. Vretenar would like to acknowledge 
the support from the Alexander von Humboldt - Stiftung.
%=========================================================================

\newpage

\end{document}